\documentclass[a4paper,11pt,floatfix]{article}
\usepackage{amsmath,mathtools,amssymb}
\usepackage{pos}
\usepackage{hyperref}
\usepackage{tabularx}
\usepackage{pgf-pie}
\usepackage{graphicx}
\newcommand{\md}{\mathrm d}
\usetikzlibrary{decorations.pathreplacing,decorations.pathmorphing}

\usepackage{siunitx}
\usepackage{soul}

\title{The thermal photon emissivity at the QCD chiral crossover from imaginary momentum correlators}
\ShortTitle{Photon emissivity at the QCD chiral crossover from imaginary momentum correlators}

\author*[a]{Ardit Krasniqi}
\author[b,c]{Marco Cè}
\author[d]{Tim Harris}
\author[e]{Renwick J. Hudspith}
\author[a,f]{\quad \quad \quad \quad Harvey B. Meyer}
\author[a]{Csaba Török}

\affiliation[a]{PRISMA$^+$ Cluster of Excellence \& Institut f\"ur Kernphysik,
Johannes Gutenberg-Universit\"at Mainz,
D-55099 Mainz, Germany}

\affiliation[b]{Dipartimento di Fisica, Università di Milano-Bicocca,
Piazza della Scienza 3, 20126 Milano, Italy}

\affiliation[c]{INFN, Sezione di Milano-Bicocca, Piazza della Scienza 3, 20126 Milano, Italy}

\affiliation[d]{Institute for Theoretical Physics, ETH Zürich,
Wolfgang-Pauli-Str. 27, 8093 Zürich, Switzerland}

\affiliation[e]{GSI Helmholtzzentrum für Schwerionenforschung, 64291 Darmstadt, Germany}

\affiliation[f]{Helmholtz Institut Mainz, Johannes Gutenberg-Universität
Mainz, Saarstr. 21, 55122 Mainz, Germany}

\emailAdd{arkrasni@uni-mainz.de}

\abstract{The thermal photon emissivity at the QCD chiral crossover is
investigated using imaginary momentum correlators. These have been
measured on a newly generated $20 \times 96^3$ lattice-QCD ensemble
with $\mathcal{O}(a)$-improved Wilson quarks and physical up, down and
strange quark masses at a temperature $T=154$\,MeV near the
pseudo-critical temperature. In order to realize the photon on-shell
condition, the spatially transverse Euclidean correlators have to be
evaluated at imaginary spatial momenta.
Employing a bounding method, we present a preliminary result on the
quantity $H_E(\omega_1)$, which corresponds to an energy-moment of the
photon spectral function $\sigma(\omega)/\omega$ defined by the weight function $1/(\omega^2+\omega_n^2)$, the $\omega_n$ being integer multiples of $2\pi T$.}

\FullConference{The 40th International Symposium on Lattice Field Theory (Lattice 2023)\\
July 31st - August 4th, 2023\\
Fermi National Accelerator Laboratory\\}

\begin{document}
\maketitle

\section{Introduction}

\begin{figure}[h!]
\center
	\includegraphics[scale=0.73]{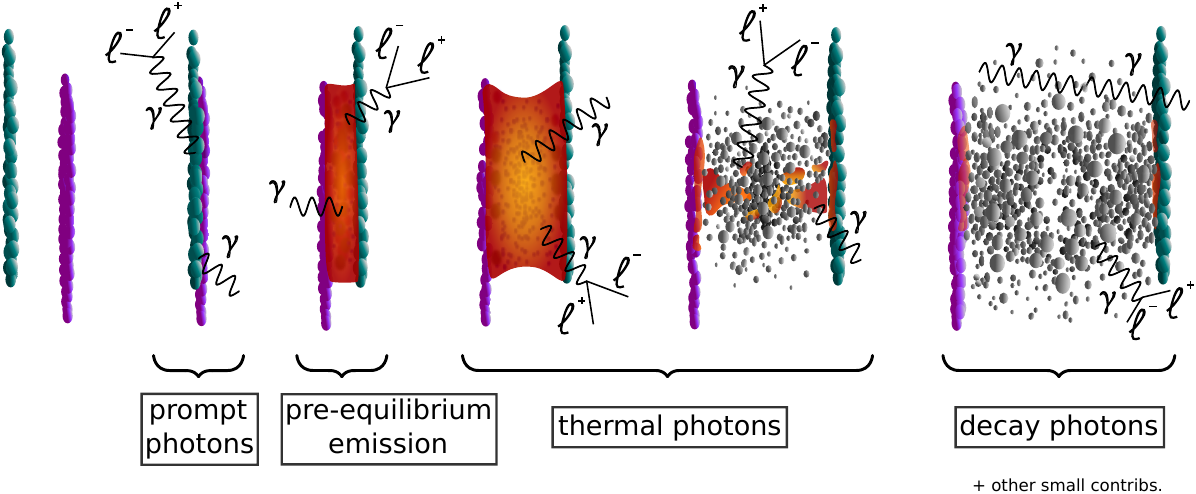}
	\caption{Sources of photons in a relativistic heavy ion collision.}
	\label{fig:photons}
\end{figure}

During a relativistic heavy-ion collision, photons can escape the quark-gluon plasma (QGP) without scattering via the strong interaction, thus  providing valuable insights into the complete space-time evolution of the QGP.

The detected photons are divided into two categories, \textbf{direct and decay photons} [see Fig.\,\ref{fig:photons}]. 
The latter result from electromagnetic decays of long-lifetime final-state hadrons, while direct photons are those produced at any stage in the collision before the final hadrons completely decouple. \textbf{Prompt photons} are real photons directly produced in hard scattering processes between partons during the initial stage of the collision, while \textbf{pre-equilibrium} photons stem from quarks and gluons that have scattered off of one another only a few times and have not yet achieved a thermal distribution, i.e. before the QGP has reached thermal equilibrium. At transverse momenta\footnote{Our transverse momentum corresponds to $\omega_1=2\pi T \approx 1$\,GeV.} $1 < p_T /\text{GeV} < 3$, one expects a sizeable contribution from \textbf{thermal photons}, both from the QGP and from the hadronic phase.

While data from the STAR experiment \cite{STAR:2016use} appear to be in agreement with theoretical models, the PHENIX and ALICE collaborations show a direct-photon excess at low transverse momenta below a few GeV \cite{Gale:2021emg, PHENIX:2022rsx, PHENIX:2014nkk, ALICE:2015xmh}, corresponding to a region of phase space dominated by thermal photons. Furthermore, these collaborations measure a photon anisotropy, $v_2$, w.r.t. the reaction plane, which is  larger than models predict \cite{PHENIX:2011oxq, PHENIX:2015igl, ALICE:2018dti}. A possible explanation put forward for the large measured anisotropy is that the photon emissivity around the crossover could be larger than assumed in the hydrodynamics-based models of the collisions, since in the later stages of the collision the photons would naturally ‘inherit’ the anisotropic flow of the strongly-interacting medium they are emitted from. 
 The inability of the majority of models to collectively account for the experimental observations is commonly denoted as the \textbf{direct photon puzzle} \cite{David:2019wpt, Gale:2021emg, Geurts:2022xmk}.

These model computations depend on theoretical estimates of the thermal photon emissivity, directly linked to the photon spectral function  $\sigma(\omega)$. Current phenomenological models have incorporated these predictions, typically derived at leading order in QCD perturbation theory \cite{Arnold:2001ba, Arnold:2001ms}. Additionally, for the low-temperature phase, relativistic kinetic theory calculations within a hot meson gas have been employed \cite{Paquet:2015lta, Turbide:2003si}. The precise method of connecting these predictions near the pseudocritical temperature $T_{pc}= 156.5(1.5)$\,MeV \cite{HotQCD:2018pds} remains uncertain, leading to the utilization of interpolations. 

Employing a bounding method, we present preliminary results on the quantity $H_E(\omega_1)$, which corresponds to the first energy-moment of the photon spectral function. $H_E(\omega_1)$ is directly accessible in lattice QCD via spatially transverse Euclidean corrrelators evaluated at imaginary spatial momentum $k=i\omega_1$ in order to realize lightlike kinematics in Euclidean space. These correlators have been measured on a newly generated  physical-mass ensemble of size $20 \times 96^3$, corresponding to a tempertaure $T=153.5(1.8)$\,MeV near the pseudocritical temperature.

\section{Numerical setup}

\begin{table}[tb]
\caption{\label{tab:E250params} Parameters and lattice spacing of the ensemble analyzed in this work.
    The lattice spacing determination is from Ref.~\cite{Bruno:2016plf}
  }
  \centering
  \begin{tabular}{c@{~~~}c@{~~~}c@{~~~}c@{~~~}c@{~~~}c}
    \hline
    \hline
    $\beta$/$a$ & $L$/$a$  & 6/$g_0^2$ & $\kappa_l$ & $\kappa_s$ & $a\,[{\rm fm}]$ \\
    \hline
    20  & 96  & 3.55  & 0.137232867 & 0.136536633  &  0.06426(76) \\
    \hline
    \hline
  \end{tabular}
  \end{table}
  
Our calculations are performed on an $N_{\mathrm{f}}=2+1$ ensemble with tree-level $\mathcal{O}(a^2)$-improved Lüscher-Weisz gauge action and non-perturbatively $\mathcal{O}(a)$-improved Wilson fermions\,\cite{Bulava:2013cta}. The ensemble has been generated using version 2.0 of the openQCD package see, Ref.\,\cite{Luscher:2012av}. 
  
We employ a single gauge ensemble of size $20\times 96^3$ of $\mathcal{O}$($a$)-improved Wilson fermions with physical quark masses at a temperature
\begin{equation}
T = \frac{1}{\beta} = \frac{1}{20a} = 153.5(1.8)\,{\rm MeV}.
\end{equation}
The physical and algorithmic parameters are listed in Tab.\,(\ref{tab:E250params}). The measurements have been performed with stochastic wall sources and a small momentum twist using the Witnesser code of Renwick J. Hudspith. 

There exists a corresponding zero-temperature \textit{Coordinated Lattice Simulations} (CLS)\,\cite{Bruno:2014jqa} ensemble with identical parameters apart from its time extent. For reference we quote the pion mass and the decay constant of this ensemble determined in Ref.\,\cite{Ce:2022kxy},
\begin{equation}
\label{eq:mpimKT0}
T=0:\qquad  m_\pi^0 = 128.1(1.3)(1.5) \,{\rm MeV}, \qquad f_\pi^0 = 87.4(0.4)(1.0)\, {\rm MeV}\,,
\end{equation}
where the first error is from the corresponding quantity in lattice units, and the second is from
the lattice spacing determination of Ref.\,\cite{Bruno:2016plf}.

\section{Preliminaries}

\subsection{The thermal photon rate}
Let $\omega$ be the energy of a photon released from a fluid cell at rest and in thermal equilibrium.
The differential photon emissivity of the QGP in leading order of the electromagnetic coupling constant $\alpha_{\text{em}}$, but to all orders of the strong coupling constant is given by \cite{McLerran:1984ay}

\begin{equation}
        \frac{d\Gamma_{\gamma}}{d\omega} = \frac{\alpha_{\text{em}}}{\pi}\frac{2\omega}{e^{\beta\omega}-1}\,\sigma(\omega) + \mathcal{O}(\alpha_{\text{em}}^2)\,,
    \end{equation}
where

\begin{equation}
        \sigma(\omega) \equiv \rho^T(\omega,k=\omega) = \frac{1}{2}\left(\delta^{ij}-\frac{k^ik^j}{k^2}\right)\rho_{ij}(\omega, \mathbf{k})
    \end{equation}
is the transverse channel spectral function. It is linked to the spatially transverse Euclidean correlator $H_E(\omega_n)$ with Matsubara frequency $\omega_n$ and imaginary spatial momentum $k=i\omega_n$ by the dispersion relation \cite{Meyer:2018xpt}
\begin{align}
    \label{eq:dispersion_relation}
    H_E(\omega_n)=-\frac{\omega_n^2}{\pi}\int_0^{\infty}\frac{\md\omega}{\omega}\frac{\sigma(\omega)}{\omega^2+\omega_n^2}\,, \ \ \ \ \omega_n=2n\pi T\,.
\end{align}

It should be stressed that computing the full energy-differential photon emissivity of a medium at thermal equilibrium from lattice QCD involves a numerically ill-posed inverse problem \cite{Ce:2022fot}.
However, energy-integrated information on the photon emissivity can be obtained without confronting an inverse problem.

\subsection{Imaginary momentum correlators}
\label{sec:imamom}

Starting from the Euclidean vector screening correlators
\begin{align}
    \label{eq:screening_vector_corr}
     G_{E,\mu\nu}(\omega_n,p_2,p_3,x_1) = \int_0^{\beta} \md x_0e^{i\omega_nx_0}\int \md^2x_{\perp} e^{i(p_2x_2+p_3x_3)} \langle J_{\mu}(x)J_{\nu}(0) \rangle\,, \quad x_{\perp} = (x_2,x_3)\,,
\end{align}
we restrict the following discussion to the transverse channel and the first Matsubara sector, defining 
\begin{align}
    \label{eq:transverse_corr}
     \notag G_{E}^T(\omega_1,p_2,x_1) &\equiv G_{E,33}(\omega_1,p_2,0,x_1) \\ &= -\int_0^{\beta} \md x_0e^{i\omega_1x_0}\int \md^2x_{\perp} e^{ip_2x_2} \langle J_3(x)J_3(0) \rangle\,,
\end{align}
as well as two special cases of Eq.\,(\ref{eq:transverse_corr}), the non-static (momentum inserted in time direction) and static (momentum inserted in spatial direction) screening correlators,
\begin{align}
    \label{eq:static_non_static_corr}
    G_{\text{ns}}^T(\omega_1,x_1) &\equiv G_E^T(\omega_1,p_2=0,x_1)\,, \\
    G_{\text{st}}^T(p,x_1) &\equiv G_E^T(\omega_1=0,p_2=p,x_1)  \,.   
\end{align}
Next, we define the Fourier transform of the non-static screening correlator,
\begin{align}
    \label{eq:fourier_transform_non_static}
    \Tilde{G}_{\text{ns}}^T(\omega_1,k) = \int_{\mathrm{R}}\md x_1 e^{ikx_1} G_{\text{ns}}^T(\omega_1,x_1)\overset{\textcolor{blue}{k=i\omega_1}}{\equiv} H_E(\omega_1)\,,
\end{align}
evaluated at Matsubara frequency $\omega_1$ and imaginary spatial momentum $k=i\omega_1$.

Making use of Eq.\,(\ref{eq:static_non_static_corr}), we finally define \cite{Meyer:2018xpt}
\begin{align}
    \label{eq:H_E_quantity}
    H_E(\omega_1) = -\int_0^{\beta}\md x_0\int \md^3x\, e^{i\omega_1x_0} e^{-\omega_1x_1} \langle J_3(x)J_3(0) \rangle <0\,.
\end{align}
Within the continuum theory, $H_E(\omega_1)$ vanishes in the vacuum, but
this property is lost at finite lattice spacing due to the lack of Lorentz symmetry. However, the property can be restored by subtracting a correlator with the same short-distance behavior and – in order to not to alter the continuum limit
–, that vanishes in the continuum. One can achieve this either by subtracting
the vacuum lattice correlator obtained at the same bare parameters or by subtracting a thermal lattice correlator having the same momentum inserted into
a spatial direction. Since the latter option does not require additional simulations at $T=0$, we proceed by subtracting the static screening correlator evaluated at momentum $p_2$ equal to the first Matsubara mode, defining the lattice subtracted correlator
\begin{align}
       \label{eq:subtracted_corr}
       H_E(\omega_1) \notag &= -\int_0^{\beta}\md x_0\int_{\mathrm{R}^3} \md^3x\, \left(e^{i\omega_1x_0}-e^{i\omega_1x_2}\right) e^{-\omega_1x_1} \langle J_3(x)J_3(0) \rangle\\
       &= 2 \int_0^{\infty} \md x_1\,\text{cosh}(\omega_1 x_1)\cdot\left[G_{\text{ns}}^T(\omega_1,x_1) - G_{\text{st}}^T(\omega_1,x_1)\right]\,.
\end{align}
We emphasize that the spatial momentum is inserted in $\hat{x}_2-$direction, i.e. transverse to the decay direction $\hat{x}_1$ as well as the direction of the Lorentz index.

The statistical precision of the screening correlators encounters two issues. First, 
the cosh-kernel multiplying the correlator difference in Eq.\,(\ref{eq:subtracted_corr}) results in an enhancement of the integrand for large $x_1$.
Furthermore, we stress that the inserted momenta are of $\mathcal{O}(1\,\text{GeV})$. Thus, we suffer from a severe signal-to-noise problem, 
resulting in exponentially amplifying uncertainties for large source-sink seperations [see l.h.s. of Fig.\,\ref{fig:model_tail_bounding_method}]. In addressing this issue, we have modeled the tail of the integrand by employing infinite-volume variants of two-state fits\footnote{This corresponds to a truncation of Eq.\,(\ref{eq: fit_static_nonstatic_corr}) for $n>2$.} of the form
\begin{align}
\label{eq:inf_volume_fit}
    G_i^T(\omega_1,x_1) = \sum_{n=1}^{2} \vert A_i^n \vert^2 e^{-m_n x_1}\,, \quad i \in \{\text{st},\text{ns}\}\,,
\end{align}
to the screening correlators. Another possibility to deal with this issue -- the bounding method -- is introduced in the next section.

\section{Bounding method}
\label{sec:bounding_method}

The screening correlators have a representation in terms of energies and amplitudes of screening states in the following form
\begin{align}
            \label{eq: fit_static_nonstatic_corr}
             G^T_{i}(\omega_r,x_1)\overset{x_1\neq 0}{=}\sum_{n=0}^{\infty}\vert{A_{i,n}^{(r)}}\vert^2 e^{-E_{i,n}^{(r)}\vert{x_1}\vert}\,, \quad i \in \{\text{st},\text{ns}\}\,, \quad \omega_r =2r\pi T\,,
\end{align}
where the index $i$ labels static and non-static correlators and $\omega_r$ denotes the $r-th$ Matsubara mode. As has been proposed for the $(g-2)$ \cite{Budapest-Marseille-Wuppertal:2017okr, RBC:2018dos, Gerardin:2019rua}, exploiting the positivity of the amplitudes and energies, we can bound the screening correlators from above and below for different values of $x_{\text{cut}}$,
\begin{align}
\begin{split}
        0 &\leq G^T_{i}(\omega_1,x_{\text{cut}})e^{-m_{\text{eff}}(x_{\text{cut}})\cdot(x_1-x_{\text{cut}})}\\ &
        \leq G^T_{i}(\omega_1,x_1)\leq G^T_{i}(\omega_1,x_{\text{cut}})e^{-E_{i,0}^{(1)}\cdot(x_1-x_{\text{cut}})}\,, \ \ x_1\geq x_{\text{cut}}\,.
\end{split}        
\end{align}
Here $m_{\text{eff}}$ denotes the effective mass of the corresponding correlator.
In contrast to the $(g-2)$, Eq.\,(\ref{eq:subtracted_corr}) involves a correlator difference. Therefore, in order to get the true upper and lower bound of $H_E$ one has to build
\begin{align}
\label{eq:bounds_for_differnce}
\left. H_E(\omega_1)\right \vert_{\text{ub}} &\propto \left. G_{\text{ns}}^T(\omega_1,x_1)\right \vert_{\text{ub}} - \left. G_{\text{st}}^T(\omega_1,x_1)\right \vert_{\text{lb}}\,,\\
            \left. H_E(\omega_1)\right \vert_{\text{lb}} &\propto \left. G_{\text{ns}}^T(\omega_1,x_1)\right \vert_{\text{lb}} - \left. G_{\text{st}}^T(\omega_1,x_1)\right \vert_{\text{ub}}\,.
\end{align}
Next, we motivate the ground state energies of both, the static, and non-static screening correlators that are used for the bounding method. Assuming two-pion ground states in both cases, for the static correlator we are dealing with two pions with back-to-back momenta,
\begin{align}
    \label{eq:energy_static}
     E_{\text{st},0}^{(1)}(p) = 2\,\sqrt{\left(\frac{p}{2}\right)^2 + m_{\pi}^2 + \left(\frac{2\pi}{L}\right)^2}\,,
\end{align}
each carrying momentum $p/2 = \pi T$ in the direction transverse to the decay direction and to the vector-index of the vector currents. The static pion-mass at zero-momentum is denoted by $m_{\pi}$ and the momentum $2\pi/L$ is in direction of the vector-index.

In the non-static sector, we have one pion at rest and the other carrying momentum equal to the first Matsubara mode in time direction, resulting in 
\begin{align}
    \label{eq:energy_non_static}
     E_{\text{ns},0}^{(1)}(\omega_1)=\sqrt{m_{\pi}^2 + \left(\frac{2\pi}{L}\right)^2} + \sqrt{E_{\pi}(\omega_1)^2 + \left(\frac{2\pi}{L}\right)^2}\,.
\end{align}

\section{Dealing with outliers}

\begin{figure}[h!]
\center
	\includegraphics[scale=0.6]{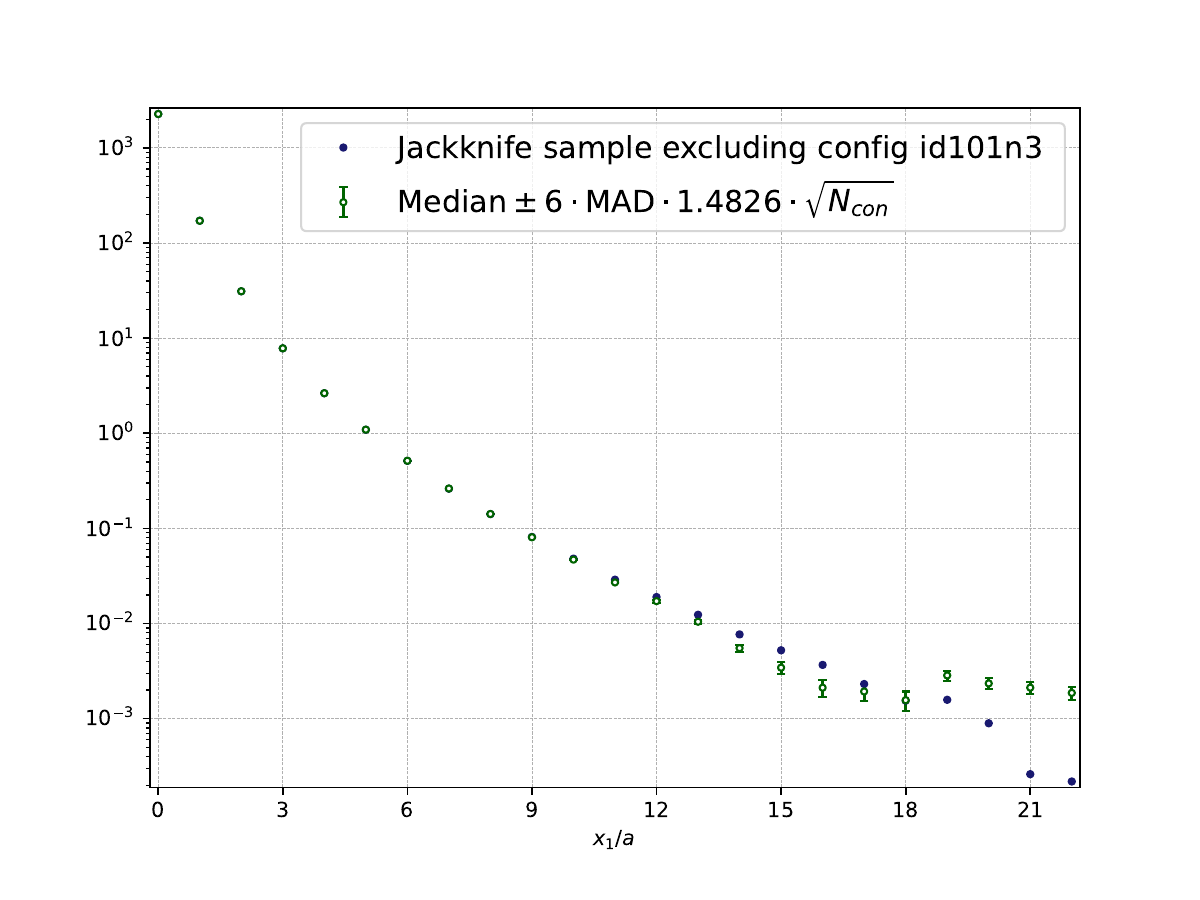}
\caption{ 
        The blue points denote the jackknife sample of the static correlator $G_{\text{st}}^T(\omega_1,x_1)$ obtained by removing one of the outlier configurations. The green points represent the median and median absolute deviation scaled by appropriate factors described in the text.}
	\label{fig:outlier}
\end{figure}

The ensemble analyzed in this work was generated in a parameter range that is difficult to access with Wilson fermions, namely physical up, down and strange quark masses right in the vicinity of the pseudocritical temperature. Therefore, additionally to the signal-to-noise problem already mentioned in Sec.\,\ref{sec:imamom}, we encounter outliers that exceed the ensemble average significantly, and accurately estimating errors becomes a challenging endeavor.

Given the sensitivity of the mean value of correlators to extreme values, we aim for a more robust statistical approach. Instead of relying on the mean and standard deviation of the sampling distribution, we utilize the robust median and median absolute deviation (MAD) to identify exceptional configurations \cite{Agadjanov:2023jha}. The latter can be related to the standard deviation via
\begin{equation}
    \label{eq:rel_sigma_mad}
    \sigma = 1.4826 \cdot \text{MAD}\,.
\end{equation}
We primarily seek significant deviations from the central position of the sampling distribution on each time slice and for every source-sink separation. Whenever we encounter a value that deviates by approximately more than $6\sigma$, we declare the configuration in question as exceptional. Following this method, we have removed five out of 1000 configurations from our analysis. In Fig.\,\ref{fig:outlier} this procedure is shown for one outlier of the static screening correlator.

Another indicator warranting careful consideration on this ensemble is the partially conserved axial current (PCAC) mass. This quantity, based on an operator identity, should be independent of the temperature. However, in lattice units we obtain \cite{Ce:2022dax}
\begin{align}
    \label{eq:PCAC_masses}
    \begin{split}
a\cdot m_{\text{PCAC}}^{vac} &= 0.001428(16)\,,\\ 
a\cdot m_{\text{PCAC}}^{N_t24} &= 0.001482(39)\,,\\
a\cdot m_{\text{PCAC}}^{N_t20} &= 0.001026(66)\,,\\
a\cdot m_{\text{PCAC}}^{N_t16} &= 0.001451(89)\,.
\end{split}
\end{align}
Therefore, the PCAC mass on this ensemble is off by a factor of $\approx 1.39$ relative to PCAC mass extracted from the corresponding vacuum box. Currently, we are initiating the production of another chain with much finer integration to address and control these issues.

\section{Results}

\begin{figure}[h!]
\center
	\includegraphics[scale=0.493]{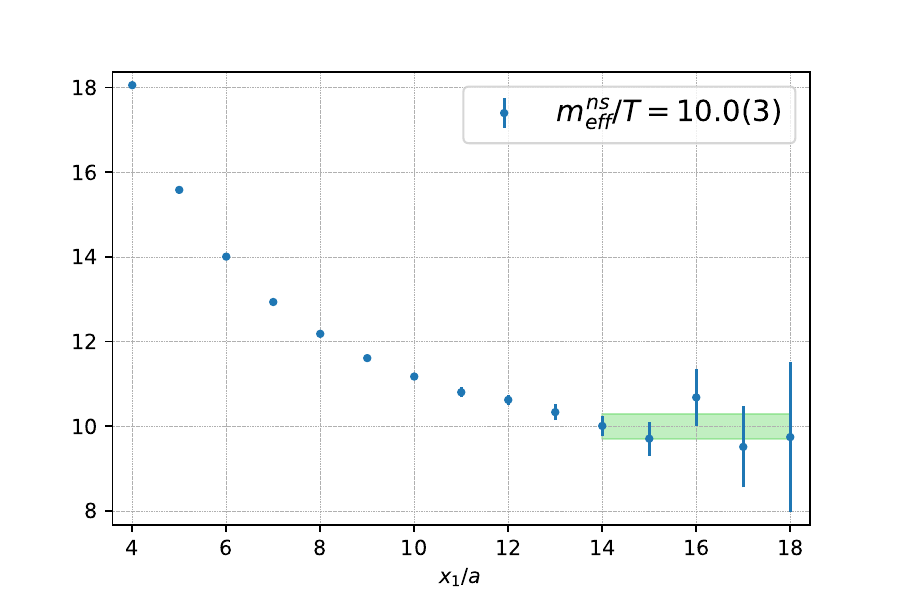}
	\includegraphics[scale=0.493]{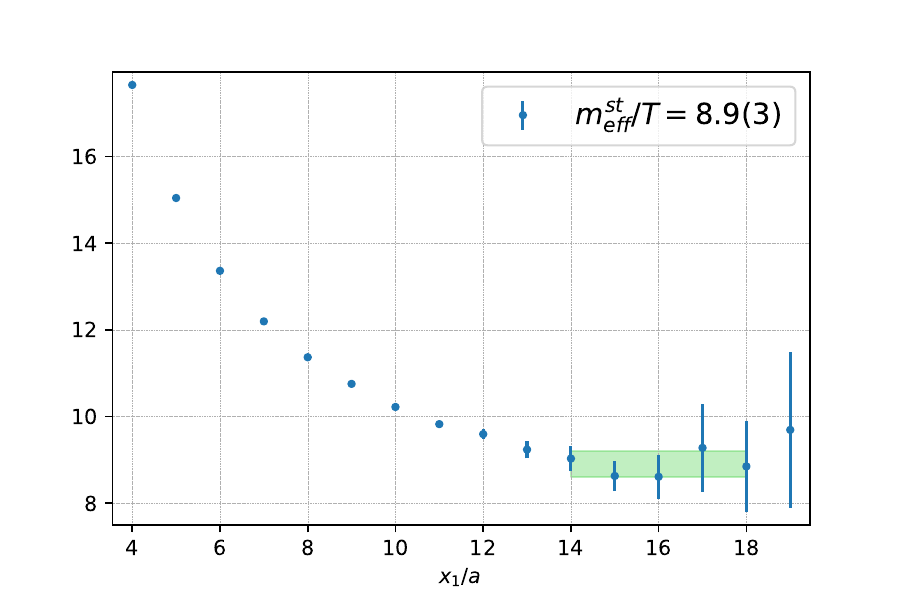}
        \caption{ 
        Effective log-masses of the non-static (left) and static (right) screening correlators.}
        \label{fig:eff_masses}
\end{figure}

\begin{figure}[h!]
\center
	\includegraphics[scale=0.44]{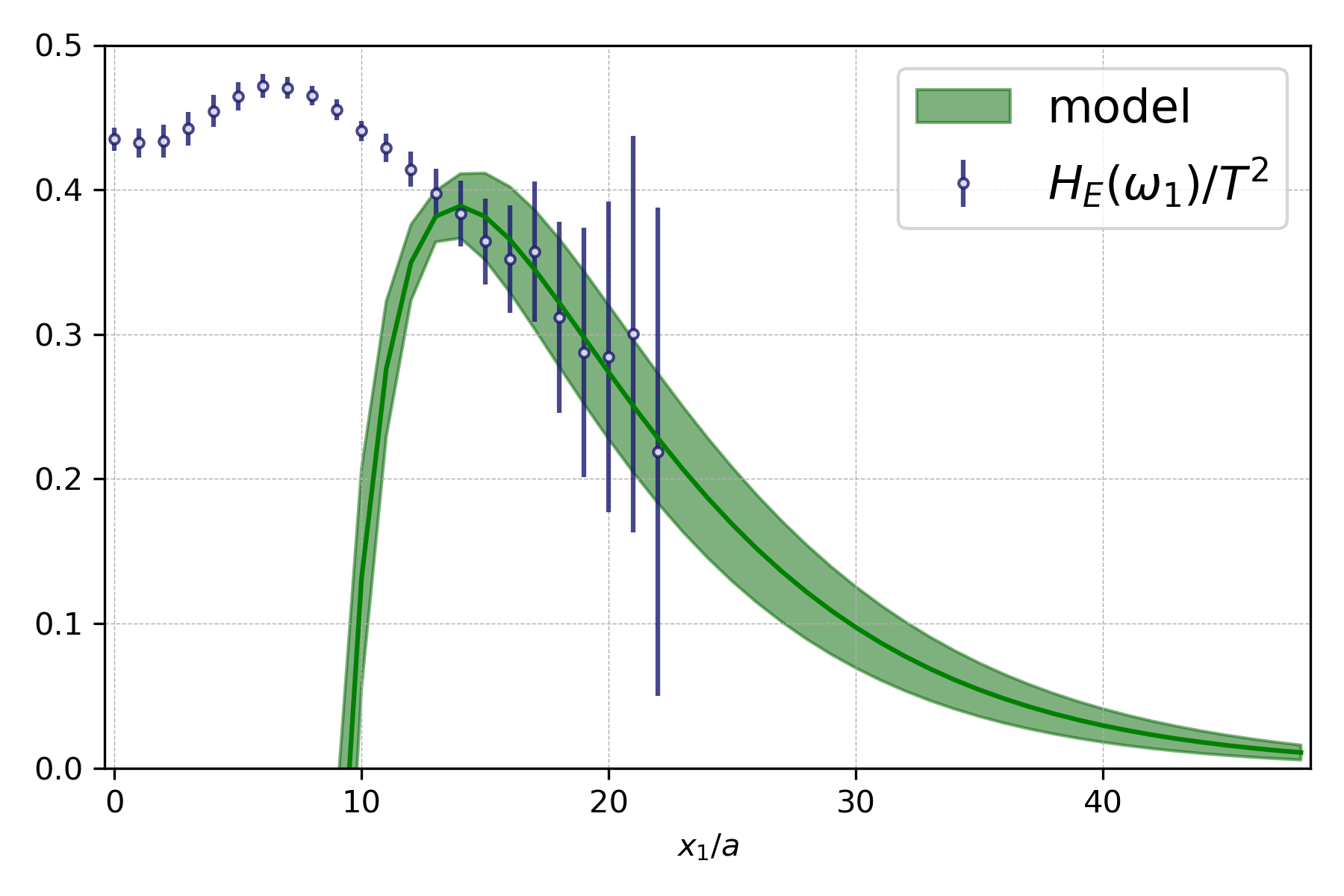}
        \includegraphics[scale=0.54]{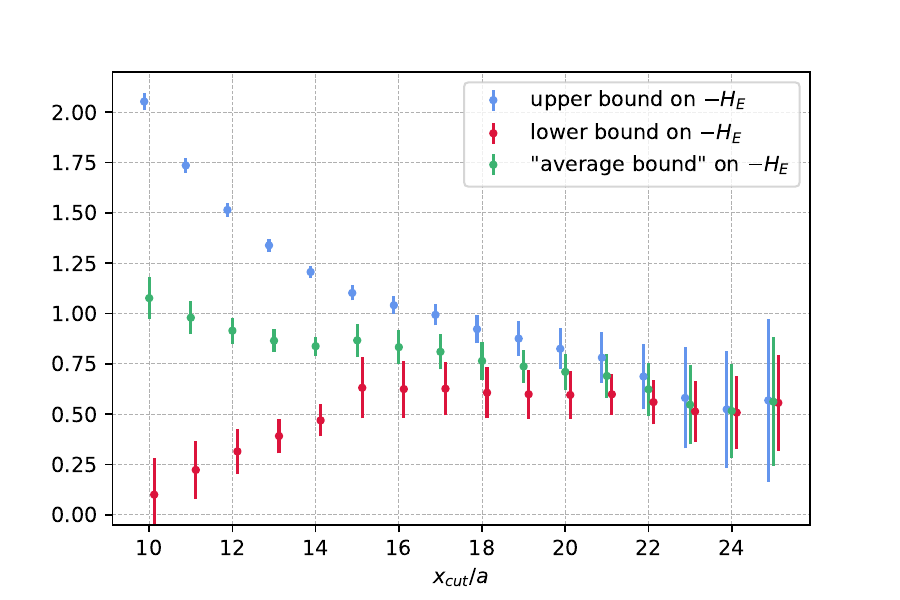}
\caption{ 
        \textbf{Left:} Plot of the $x_1$-symmetrized integrand of Eq.\,(\ref{eq:subtracted_corr}) together with the result from modelling the tail. \textbf{Right:} Lower and upper bounds on $H_E(\omega_1)$ that converge toward that quantity for $x_{cut} \rightarrow \infty$.}
	\label{fig:model_tail_bounding_method}
\end{figure}

Finally, we present our preliminary results. As discussed in Secs.\,\ref{sec:imamom} and \ref{sec:bounding_method}, the effective masses of the screening correlators are pivotal. They play a crucial role in estimating the lower bound in the bounding method and are essential for modelling the tails of the correlators. In Fig.\,\ref{fig:eff_masses}, the l.h.s. illustrates the effective mass in the non-static sector, while the r.h.s. depicts the effective mass of the static sector. Expressed in temperature units we obtain,
\begin{align}
    \label{eq:eff_masses}
    E_{\text{ns},0}^{(1)}/T \equiv m_{\text{eff}}^{\text{ns}}/T &= 10.0(3)\,, \\
     E_{\text{st},0}^{(1)}/T \equiv m_{\text{eff}}^{\text{st}}/T &= 8.9(3)\,.
\end{align}
From Eq.\,(\ref{eq:dispersion_relation}) we know that $H_E(\omega_1)$ is a measure of the integrated photon emissivity. The l.h.s. of Figure\,\ref{fig:model_tail_bounding_method} shows the integrand of Eq.\,(\ref{eq:subtracted_corr})
together with the result from modelling the tail. As a result of the exponential signal-to-noise problem, we start to loose the signal for source-sink seperations greater than $x_1/a\approx 20$. The short-distance contribution of $H_E(\omega_1)$ is determined using the trapezoidal formula, while the long-distance contribution is integrated analytically using the fit parameters of the infinite volume fit
ansatz given in Eq.\,(\ref{eq:inf_volume_fit}). The transition to the modeled tail has been seamlessly incorporated by employing a smooth step function \cite{Ce:2023oak}. Opting for different switching points introduces a systematic error that is negligible in comparison to the substantial statistical error. We quote
\begin{equation}
    \label{eq:result_tail}
    \left.\frac{H_E(\omega_1)}{T^2}\right\vert_{\text{model}} = -0.54(5)\,,
\end{equation}
as our preliminary result. 

A second estimator can be obtained using the bounding method. The result is shown in the right panel of Fig.\,\ref{fig:model_tail_bounding_method}. The correlators are bounded from below with the aid of the effective masses and from above with the energies assuming two-pion
ground states. For $x_{cut} \rightarrow \infty$ lower and upper bound should converge to the true value of $H_E(\omega_1)$. As a preliminary result, we quote the average value of upper and lower bound,
\begin{equation}
    \label{eq:result_bounding_method}
    \left.\frac{H_E(\omega_1)}{T^2}\right\vert_{\text{bound}} = -0.62(13)\,,
\end{equation}
obtained for $x_{cut}/a=22$.

\section{Conclusion}
To assess the collective characteristics of the hadronic phase as the temperature rises, commonly the \textit{hadron resonance gas} (HRG) model \cite{Hagedorn:1984hz} is employed. This model characterizes the thermodynamic properties of the system through the summation of individual contributions from non-interacting hadron species up to a specific cut-off mass.
Comparing our lattice estimate of the quark number susceptibility (QNS) $\chi_q/T^2 = 0.73(1)$ to the value obtained using the HRG-model $\left. \chi_q/T^2\right\vert_{\text{HRG}} = 0.70$ suggests the assumtion that the QNS can still be described approximately using hadronic degress of freedom.

However, in temperature units the $H_E$-quantity near the pseudocritical temperature is in the same ballpark as our result \cite{Torok:2022vki, Ce:2023oak}
\begin{equation}
    \label{eq:result_bounding_method}
    \left.\frac{H_E(\omega_1)}{T^2}\right\vert_{\text{high T}} = -0.670(6)\,,
\end{equation}
in the high-temperature phase at $T\approx 250$\,MeV ($\chi_q/T^2 = 0.88(1)$), despite there being fewer charge degrees of freedom around the crossover. 

\acknowledgments{This work was supported by the European Research Council (ERC) under the European
Union’s Horizon 2020 research and innovation program through Grant Agreement
No.\ 771971-SIMDAMA, as well as by the Deutsche Forschungsgemeinschaft 
(DFG, German Research Foundation) through the Cluster of Excellence “Precision Physics,
Fundamental Interactions and Structure of Matter” (PRISMA+ EXC 2118/1) funded by
the DFG within the German Excellence strategy (Project ID 39083149).
The research of M.C. is funded through the MUR program for young researchers ``Rita Levi Montalcini''.
The generation of gauge configurations of the $20 \times 96^3$ ensemble was performed at Forschungszentrum J\"ulich allocated under SIMCHIC.
The measurements of the two-point functions have been partly performed on our local machine MogonII and partly on  Noctua2 in Paderborn as part of a small-time NHR-proposal. For the estimation of the statistical errors of our (derived) observables we use the $\Gamma$ method in the implementation of the pyerrors package introduced in Ref.\,\cite{Joswig:2022qfe}.
\newpage

\bibliography{he2p1.bib}

\end{document}